\documentclass[conference]{IEEEtran}
\usepackage{amsmath, graphicx, amsfonts, amssymb, amsthm, float, cite}
\usepackage[top=0.85in, bottom=0.9in, left=0.65in, right=0.65in]{geometry}
\newtheorem{theorem}{Theorem}

\ifCLASSINFOpdf
\else
\fi

\begin{document}
\bibliographystyle{IEEEtran}
\title{Detection in Analog Sensor Networks \\
with a Large Scale Antenna Fusion Center}

\author{\IEEEauthorblockN{Feng Jiang, Jie Chen, and A. Lee Swindlehurst}
\IEEEauthorblockA{Center for Pervasive Communications and Computing\\University of California at Irvine\\Irvine, CA, 92697, USA\\Email:\{feng.jiang, jie.chen, swindle\}@uci.edu\\}}


%


\maketitle


%

\begin{abstract}
We consider the distributed detection of a zero-mean Gaussian signal in an analog wireless sensor network with a fusion center (FC) configured with a large number of antennas.  The transmission gains of the sensor nodes are optimized by minimizing the ratio of the log probability of detection (PD) and log probability of false alarm (PFA). We show that the problem is convex with respect to the squared norm of the transmission gains, and that a closed-form solution can be found using the Karush-Kuhn-Tucker conditions.  Our results indicate that a constant PD can be maintained with decreasing sensor transmit gain provided that the number of antennas increases at the same rate.  This is contrasted with the case of a single-antenna FC, where PD is monotonically decreasing with transmit gain.  On the other hand, we show that when the transmit power is high, the single- and multi-antenna FC both asymptotically achieve the same PD upper bound.
\end{abstract}

\section{Introduction}
Signal detection and parameter estimation in wireless sensor networks (WSN) have been widely studied \cite{Chamberland:2004, Cui:2007, Quan:2010, Jiang:2013}, and much of the existing work has assumed a fusion center (FC) equipped with a single antenna. It is well known that multiple antennas can effectively increase the system capacity of wireless links, and recent work has investigated the benefit provided by multiple antennas in WSN detection and estimation problems \cite{Smith:2009, Banavar:2012, Jiang:20122}. In \cite{Smith:2009, Banavar:2012}, power allocation problems were formulated with a multi-antenna FC under Rayleigh fading channels, and it was shown that when the number of sensors is large, the relative performance gain of a multi-antenna FC over a single-antenna FC is bounded by constants unrelated to the number of antennas. In \cite{Jiang:20122}, a phase-shift-and-forward method was proposed, and the results show that in some cases the estimation variance can be reduced by a factor proportional to the number of antennas.  Recent research in cellular communication systems has shown that employing a base station with a massive number of antennas has considerable advantages, including the ability to achieve a constant signal-to-interference ratio with arbitrarily small transmit powers at the single-antenna terminals \cite{Marzetta:2010,Hoydis:2013,Ngo:2013}.

In this paper, we consider an analog WSN and we assume a fusion center with a massive number of antennas.  In our model, the sensor nodes measure a random signal of interest corrupted by measurement noise. The noisy measurements are amplified and forwarded to the FC over a coherent multiple access channel, and based on the received signal, the FC uses the Neyman-Pearson (NP) rule to decide whether or not the signal of interest is present. We optimize the detection performance of the FC by adjusting the transmission gains of the sensors by minimizing the ratio of the log PD and log PFA.  We show that the resulting optimization problem is convex and that a closed-form solution can be found using the Karush-Kuhn-Tucker (KKT) conditions. We also derive performance bounds and investigate the conditions under which the benefit of multiple antennas can be exploited.

\section{Signal Model}
We consider a general Gaussian detection problem, in which a zero-mean Gaussian signal of interest $\theta$ is observed by a collection of single-antenna sensors in the presence of zero-mean Gaussian noise.  Each sensor coherently amplifies and forwards its measurement to an FC that possess $M$ antennas, and a decision is made at the FC as to the presence/absence of the signal.  The sensor measurement model under the two hypotheses is
\begin{eqnarray}
\mathcal{H}_0: s_i\!\!\!\!&=&\!\!\!v_i\nonumber\\
\mathcal{H}_1: s_i\!\!\!\!&=&\!\!\!\theta+v_i\;,\nonumber
\end{eqnarray}
where the measurement noise $v_i$ has distribution $\mathcal{CN}(0,\sigma_{v,i}^2)$, and $\theta$ is distributed as $\mathcal{CN}(0,\sigma_{\theta}^2)$. The sensor node first multiplies the measurement $s_i$ by a complex gain $a_i$ and then forwards the product to the FC through a wireless fading channel. The received signal at the FC is
\begin{eqnarray}\label{eq:rsig}
\mathcal{H}_0: \mathbf{y}\!\!\!\!&=&\!\!\!\mathbf{H}\mathbf{D}\mathbf{v}+\mathbf{n}\\
\mathcal{H}_1: \mathbf{y}\!\!\!\!&=&\!\!\!\mathbf{H}\mathbf{a}\theta+\mathbf{H}\mathbf{D}\mathbf{v}+\mathbf{n}\;\nonumber,
\end{eqnarray}
where $\mathbf{H}=[\mathbf{h}_1,\cdots,\mathbf{h}_N]$ and $\mathbf{h}_i\in\mathbb{C}^{M\times 1}$ is the channel gain between the $i$th sensor and the FC, $\mathbf{a}=[a_1,\cdots, a_N]^T$ contains the transmission gains and $(\cdot)^T$ denotes the transpose, $\mathbf{D}=\mathrm{diag}\{a_1,\cdots, a_N\}$, the measurement noise vector $\mathbf{v}=[v_{1},\cdots, v_N]^T$ has covariance $\mathbf{V}=\mathrm{diag}\{\sigma_{v,1}^2,\cdots,\sigma_{v,N}^2\}$, and $\mathbf{n}$ is additive Gaussian noise at the FC with distribution $\mathcal{CN}(0,\sigma^2_n\mathbf{I}_{M})$, where $\mathbf{I}_M$ is the $M\times M$ identity matrix.

We assume the FC uses the NP criterion to distinguish between the hypotheses $\mathcal{H}_0$ and $\mathcal{H}_1$. The NP detector decides $\mathcal{H}_1$ if \cite{Kay:19932}
\begin{equation}\label{eq:lr}
L(\mathbf{y})=\frac{p(\mathbf{y};\mathcal{H}_1)}{p(\mathbf{y};\mathcal{H}_0)}>\gamma\;,
\end{equation}
where $\gamma$ is a predefined threshold, and $p(\mathbf{y};\mathcal{H}_k)$ and is the conditional probability density function (PDF) of $\mathbf{y}$ under $\mathcal{H}_k$. Since $\mathbf{y}$ is Gaussian distributed under either hypothesis, we have
\begin{eqnarray}
p(\mathbf{y};\mathcal{H}_1)\!\!&\!\!=\!\!\!&\!\!\frac{1}{\pi^M\mathrm{det}(\mathbf{C}_s+\mathbf{C}_w)}\mathrm{exp}\left(-\mathbf{y}^H(\mathbf{C}_s+\mathbf{C}_w)^{-1}\mathbf{y}\right)\nonumber\\
p(\mathbf{y};\mathcal{H}_0)\!\!&\!\!=\!\!\!&\!\!\frac{1}{\pi^M\mathrm{det}(\mathbf{C}_w)}\mathrm{exp}\left(-\mathbf{y}^H\mathbf{C}_w^{-1}\mathbf{y}\right)\;\nonumber,
\end{eqnarray}
where $(\cdot)^H$ is the conjugate transpose, $\mathbf{C}_w=\mathbf{H}\mathbf{D}\mathbf{V}\mathbf{D}^{H}\mathbf{H}^{H}+\sigma^2_n\mathbf{I}_M$ and $\mathbf{C}_s=\sigma_{\theta}^2\mathbf{H}\mathbf{a}\mathbf{a}^H\mathbf{H}^H$. Thus, after plugging $p(\mathbf{y};\mathcal{H}_1)$ and $p(\mathbf{y};\mathcal{H}_0)$ into (\ref{eq:lr}), we have the following test statistic
\begin{equation}
\sigma_{\theta}^2\mathbf{y}^H\mathbf{C}_w^{-1}\mathbf{H}\mathbf{a}\mathbf{a}^H\mathbf{H}^H\mathbf{C}_w^{-1}\mathbf{y}>\gamma^{'}\;,\nonumber
\end{equation}
where $\gamma^{'}=\ln\left(\gamma(1+\sigma_{\theta}^2g(\mathbf{a}))\right)(1+\sigma_{\theta}^2g(\mathbf{a}))$ and $g(\mathbf{a})=\mathbf{a}^H\mathbf{H}^H\mathbf{C}_w^{-1}\mathbf{H}\mathbf{a}$.
The probability of detection $P_{D}$ and probability of false alarm $P_{FA}$ are defined as
\begin{eqnarray}
P_{D}\!\!\!\!\!&=&\!\!\!\!\mathrm{Pr}\left\{\sigma_{\theta}^2\mathbf{y}^H\mathbf{C}_w^{-1}\mathbf{H}\mathbf{a}\mathbf{a}^H\mathbf{H}^H\mathbf{C}_w^{-1}\mathbf{y}>\gamma^{'}|\mathcal{H}_1\right\}\nonumber\\
P_{FA}\!\!\!\!\!&=&\!\!\!\!\mathrm{Pr}\left\{\sigma_{\theta}^2\mathbf{y}^H\mathbf{C}_w^{-1}\mathbf{H}
\mathbf{a}\mathbf{a}^H\mathbf{H}^H\mathbf{C}_w^{-1}\mathbf{y}>\gamma^{'}|\mathcal{H}_0\right\}\; ,\nonumber
\end{eqnarray}
and are calculated to be
\begin{eqnarray}\label{eq:pd2}
P_{D}\!\!\!\!\!&=&\!\!\!\!\exp\left(-\frac{\gamma^{'}}{{\sigma_{\theta}^4g(\mathbf{a})^2+
\sigma_{\theta}^2}g(\mathbf{a})}\right) \nonumber\\
P_{FA}\!\!\!\!\!&=&\!\!\!\!\mathrm{exp}\left(-\frac{\gamma^{'}}{\sigma_{\theta}^2g(\mathbf{a})}\right)\; .\label{eq:pf}
\end{eqnarray}
The goal is to choose a suitable value for sensor transmission gains $\mathbf{a}$ in order to achieve good $P_D$ and $P_{FA}$ performance.

For NP decision rules, one typically maximizes $P_{D}$ assuming a given constraint on $P_{FA}$.  However, according to (\ref{eq:pf}), requiring $P_{FA}\le\epsilon$ is equivalent to
\begin{equation}
\ln\left(\gamma(1+\sigma_{\theta}^2g(\mathbf{a}))\right)\left(1+\frac{1}{\sigma_{\theta}^2g(\mathbf{a})}\right)\ge -\ln(\epsilon)\;,\nonumber
\end{equation}
which leads to an intractable optimization with respect to $\mathbf{a}$.  In this paper, we take a different approach and attempt to minimize the ratio $\frac{\ln{P_{D}}}{\ln{P_{FA}}}$, which is given by
\begin{equation}\label{eq:dr}
\frac{\ln{P_{D}}}{\ln{P_{FA}}}=\frac{1}{1+\sigma_{\theta}^2g(\mathbf{a})}\; ,
\end{equation}
implying that $g(\mathbf{a})$ must be maximized.  According to (\ref{eq:pf}), the threshold required to achieve $P_{FA}=\epsilon$ is
\begin{equation}
\gamma'=-\sigma_\theta^2 g(\mathbf{a}) \ln{\epsilon}\;.\nonumber
\end{equation}

\section{Problem Formulation and Solution}
Under a sum constraint on the transmission gains of the sensors, minimizing the ratio $\frac{\ln{P_{D}}}{\ln{P_{FA}}}$ is equivalent to
\begin{eqnarray}
\max_{\mathbf{a}} && g(\mathbf{a})\label{eq:opt}\\
s.t. && \mathbf{a}^{H}\mathbf{a}=P\;,\nonumber
\end{eqnarray}
where $g(\mathbf{a})=\mathbf{a}^H\mathbf{H}^H(\mathbf{H}\mathbf{D}\mathbf{V}\mathbf{D}^H\mathbf{H}^H+\sigma_n^2\mathbf{I}_M)^{-1}\mathbf{H}\mathbf{a}\;$ and $P$ denotes the gain constraint.
For our analysis, we model the wireless fading channel between sensor node $i$ and the FC as
\begin{equation}\label{eq:channel}
\mathbf{h}_i=\frac{\tilde{\mathbf{h}}_i}{d_i^{\alpha}}\;,
\end{equation}
where $d_i$ is the distance between the sensor and FC, $\alpha$ is the path loss exponent, and $\mathbf{\tilde{h}}_i\in\mathbb{C}^{M\times 1}$ is a complex Gaussian vector with distribution $\mathcal{CN}(0,\mathbf{I}_M)$.  Using this channel model, our main result regarding problem (\ref{eq:opt}) is summarized below as Theorem~1.

\begin{theorem}
Assuming Rayleigh fading wireless channels as in~(\ref{eq:channel}), as the number of FC antennas $M$ tends to infinity, the transmit gain $|a_i|^2$ at each sensor can be reduced by $1/M$ to achieve the same optimal $P_D$ for a given fixed $P_{FA}$.
\end{theorem}
\begin{IEEEproof}
We will show that as $M\rightarrow \infty$, the function $g(\mathbf{a})$ in~\eqref{eq:dr} and~\eqref{eq:opt} remains constant if the product $M |a_i|^2$ is held constant.  Thus an increase in $M$ allows for a decrease in $|a_i|^2$ by $M$ to achieve the same performance.  We first use the matrix inversion lemma to show that
\begin{align}\label{eq:inv}
&\left(\mathbf{H}\mathbf{D}\mathbf{V}\mathbf{D}^H\mathbf{H}^H+\sigma^2_n\mathbf{I}_M\right)^{-1} \nonumber\\
=&\;\frac{1}{\sigma_n^2}\mathbf{I}_M-\frac{1}{\sigma_n^4}\mathbf{H}\left(\mathbf{E}^{-1}+\frac{1}{\sigma^2_n}\mathbf{H}^H\mathbf{H}\right)^{-1}\mathbf{H}^H\;,
\end{align}
where $\mathbf{E}=\mathbf{D}\mathbf{V}\mathbf{D}^H$. Note that in the above derivation, we assume that the norm $|a_i|>0$ to guarantee the matrix inverse $\mathbf{E}^{-1}$ exists. Substituting (\ref{eq:inv}) into $g(\mathbf{a})$ yields
\begin{eqnarray}\label{eq:inv2}
g(\mathbf{a})\!\!\!\!&=&\!\!\!\!\frac{1}{\sigma_n^2}\mathbf{a}^H\mathbf{H}^H\mathbf{H}\mathbf{a}\nonumber\\
&&-\frac{1}{\sigma_n^4}\mathbf{a}^H\mathbf{H}^H\!\mathbf{H}\!\left(\!\mathbf{E}^{-1}\!\!+\!\!\frac{1}{\sigma^2_n}\mathbf{H}^H\mathbf{H}\!\right)^{\!\!\!-1}\!\!\!\mathbf{H}^H\mathbf{H}\mathbf{a}\;.
\end{eqnarray}
For large $M$, the product $\mathbf{H}^H\mathbf{H}$ converges almost surely as follows:
\begin{equation}\label{eq:approx}
\lim_{M\to\infty}\frac{1}{M}\mathbf{H}^H\mathbf{H}= \mathrm{diag}\left\{\frac{1}{d_1^{2\alpha}},\cdots, \frac{1}{d_N^{2\alpha}}\right\}\; ,
\end{equation}
and substituting (\ref{eq:approx}) into (\ref{eq:inv2}) yields, after some calculations,
\begin{equation}\label{eq:approx2}
\lim_{M\to\infty} g(\mathbf{a})=\lim_{M\to\infty}\sum_{i=1}^N\frac{M|a_i|^2}{\sigma_n^2d_i^{2\alpha}+M|a_i|^2\sigma_{v,i}^2}\;.
\end{equation}
We see that $g(\mathbf{a})$ remains asymptotically unchanged as long as the product $M |a_i|^2$ is held constant, and thus asymptotically equivalent detection performance can be achieved if any decrease in sensor transmit power is accompanied by a corresponding increase in the number of FC antennas.
\end{IEEEproof}

Based on (\ref{eq:approx2}), when $M \rightarrow \infty$, the original problem (\ref{eq:opt}) can be rewritten as
\begin{eqnarray}\label{eq:opt2}
\max_{|a_i|^2} &&\sum_{i=1}^N\frac{M|a_i|^2}{\sigma_n^2d_i^{2\alpha}+M|a_i|^2\sigma_{v,i}^2}\\
 s.t. &&\sum_{i=1}^N |a_i|^2=P\;\label{eq:c2}.     \nonumber
 \end{eqnarray}
Although $|a_i|$ should be positive to make (\ref{eq:inv}) valid, in problem (\ref{eq:opt2}) we allow $|a_i|=0$. If $|a_i|=0$, sensor $i$ will not transmit and the $|a_i|$ will not appear in~(\ref{eq:inv}). Define a new variable $x_i=|a_i|^2$, so that problem (\ref{eq:opt2}) is equivalent to
\begin{eqnarray}\label{eq:opt3}
\min_{x_i} &&\sum_{i=1}^N\frac{-Mx_i}{\sigma_n^2d_i^{2\alpha}+M\sigma_{v,i}^2x_i}\\
 s.t. && \sum_{i=1}^N x_i=P\nonumber\\
      && 0\le x_i \;\nonumber.
\end{eqnarray}
In problem (\ref{eq:opt3}), the objective function is the sum of $N$ convex functions of variable $x_i$, and the constraints are linear with respect to $x_i$, so problem (\ref{eq:opt3}) is convex and we can find the solution using the KKT conditions \cite{Boyd:2004}.  The optimal solution to~(\ref{eq:opt}) is given by
\begin{equation}\label{eq:optsol}
|a_{i}^*|=\sqrt{\frac{\left(\sqrt{\frac{M\sigma_n^2d_i^{2\alpha}}{\lambda}}-\sigma_n^2d_i^{2\alpha}\right)^{+}}{M\sigma_{v,i}^2}}\;,
\end{equation}
where $(x)^{+}=\max(0,x)$ and $\lambda$ is a positive constant chosen such that $\sum_{i=1}^N |a_i^{*}|^2=P$. Due to space limitations, the derivation of~\eqref{eq:optsol} is omitted and the details can be found in \cite{Jiang:20132}.

\section{Detection Performance Analysis}
\subsection{High Transmit Power}
From (\ref{eq:approx2}), it is clear that for very large $M$, $g(\mathbf{a})$ is upper bounded by
\begin{equation}\label{eq:ub}
g(\mathbf{a})\le\sum_{i=1}^N\frac{1}{\sigma_{v,i}^2}\;.
\end{equation}
When $P\to\infty$, the upper bound in (\ref{eq:ub}) can be asymptotically achieved even with an equal power allocation $|a_i|=P/N$. Also, we see that to maximize the upper bound for $g(\mathbf{a})$, all the sensors should transmit. Plugging (\ref{eq:ub}) into (\ref{eq:dr}), we have the following upper bound for $P_D$:
\begin{equation}\label{eq:ubm}
P_{D}\le P_{FA}^{\frac{1}{1+\sigma_{\theta}^2\sum_{i=1}^N\frac{1}{\sigma_{v,i}^2}}}\;,\;P\to\infty\;.
\end{equation}

\subsection{Low Transmit Power}
To find a lower bound when the transmit power is small, we first assume the following suboptimal choice for the transmission gains: $|\bar{a}_{i}|=\sqrt{\frac{\sigma_n^2 d_i^{2\alpha}}{2M\sigma_{v,i}^2}}$, which will result in
\begin{equation}
\label{eq:Psub}
P=\sum_{i=1}^N|\bar{a}_i|^2=\frac{1}{2M}\sum_{i=1}^{N}\frac{\sigma_n^2 d_i^{2\alpha}}{\sigma_{v,i}^2}\;,
\end{equation}
and hence $P\rightarrow 0$ as $M\rightarrow \infty$. Plugging $|\bar{a}_{i}|=\sqrt{\frac{\sigma_n^2 d_i^{2\alpha}}{2M\sigma_{v,i}^2}}$ into equation (\ref{eq:approx2}), we have
\begin{equation}
g(\mathbf{\bar{a}})=\frac{1}{3}\sum_{i=1}^N\frac{1}{\sigma_{v,i}^2}\;,\nonumber
\end{equation}
which can serve as a lower bound for $g(\mathbf{a})$ when evaluated at the optimal solution $\mathbf{a}^*$ obtained using~(\ref{eq:optsol}) and the value of $P$ in~(\ref{eq:Psub}) as the power constraint:
\begin{equation}\label{eq:glb}
g(\mathbf{a}^{*})\ge \frac{1}{3}\sum_{i=1}^N\frac{1}{\sigma_{v,i}^2}\;.
\end{equation}
Note that the lower bound in (\ref{eq:glb}) is one third of the upper bound in (\ref{eq:ub}). Substituting (\ref{eq:glb}) into (\ref{eq:dr}), we have
\begin{equation}\label{eq:lbm}
P_{D}\ge P_{FA}^{\frac{1}{1+\frac{\sigma_{\theta}^2}{3}\sum_{i=1}^N\frac{1}{\sigma_{v,i}^2}}}\;,\; P\to 0\;.
\end{equation}

\subsection{Single-antenna FC}
For comparison, we also investigate the detection performance of a single-antenna FC. In this case, the received signal in (\ref{eq:rsig}) reduces to
\begin{eqnarray}
\mathcal{H}_0: y\!\!\!\!&=&\!\!\!\!\mathbf{a}^H\mathbf{F}\mathbf{v}+n\nonumber\\
\mathcal{H}_1: y\!\!\!&=&\!\!\!\mathbf{a}^H\mathbf{h}\theta+\mathbf{a}^H\mathbf{F}\mathbf{v}+n\;,\nonumber
\end{eqnarray}
where $\mathbf{F}=\mathrm{diag}\{h_1,\cdots,h_N\}$, $h_i$ denotes the wireless channel between the $i$th sensor and the FC, $n$ is the additive noise at the FC with distribution $\mathcal{CN}(0,\sigma^2_n)$ and $\mathbf{h}=[h_1,\cdots,h_N]^T$.  Similar to the multi-antenna analysis, for the single-antenna FC, the optimal solution that minimizes the ratio $\frac{\ln {P_D^s}}{\ln P_{FA}^s}$ under the sum gain constraint $P$ is
\begin{equation}\label{eq:singlefc}
\tilde{\mathbf{a}}^{*}=\sqrt{\frac{P}{\mathbf{h}^H\mathbf{B}^{-2}\mathbf{h}}}\mathbf{B}^{-1}\mathbf{h}\;,
\end{equation}
where $\mathbf{B}=\mathbf{F}\mathbf{V}\mathbf{F}^H+\frac{\sigma^2_n}{P}\mathbf{I}_N$.  Based on (\ref{eq:singlefc}) we have the following bounds:
\begin{eqnarray}
P_{D}^s\!\!\!\!&\le&\!\!\!\!{P_{FA}^s}^{\frac{1}{1+\sigma_{\theta}^2\sum_{i=1}^N\frac{1}{\sigma_{v,i}^2}}}\;, \; P\to\infty\;\label{eq:ubsh}\\
P_{D}^s\!\!\!\!&\le&\!\!\!\!{P_{FA}^s}^{\frac{1}{1+\frac{\sigma_{\theta}^2P}{\sigma_n^2}\mathbf{h}^H\mathbf{h}}}\;,\quad\; \;\;\!P \to 0\label{eq:ubsl}\;.
\end{eqnarray}
Note that these bounds are tight for the limiting values of $P$\;. Comparing (\ref{eq:ubm}) and (\ref{eq:ubsh}), we observe that when $P\to\infty$, the single- and multi-antenna FCs asymptotically achieve the same detection performance.  However, when $P\to 0$, $P_{D}^s$ converges to $P_{FA}^s$ in the single-antenna case, while $P_D$ is lower bounded by a constant strictly larger than $P_{FA}$ in the multi-antenna case. In the next section, we will present several numerical results to verify these conclusions.

\section{Simulation Results}\label{sec:three}
In the simulations presented here, we assume $\sigma_{\theta}^2=1$ and $N=10$ sensors. The distances $d_i$ are uniformly distributed on $[2, 10]$, and the path loss exponent $\alpha$ is set to $1$. The power of the additive noise at the FC is set to $\sigma_n^2=0.3$ and the measurement noise powers $\sigma_{v,i}^2$ are uniformly distributed over $[0.25, 0.5]$. We fix $P_{FA} = P_{FA}^s=0.05$ and compare the probability of detection $P_{D}$ and $P_{D}^s$.  For each channel realization ($\mathbf{H}$ or $\mathbf{h}$), $10000$ detection tests are carried out for different signal and noise realizations, and each point in the plots is obtained by averaging over $300$ channel realizations.

\begin{figure}[!htb]
\centering
\includegraphics[height=2.5in, width=3.3in]{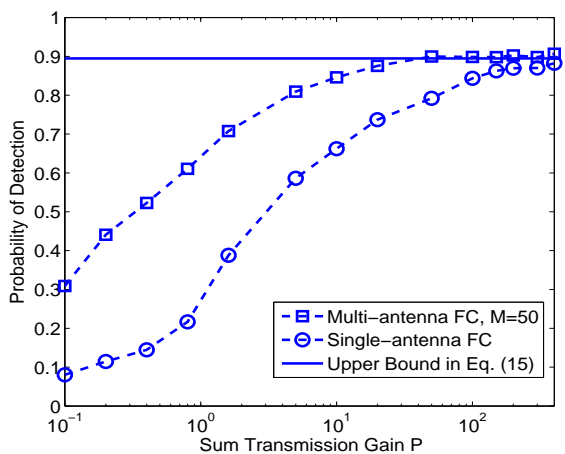}
\caption{Probability of detection vs. the value of $P$.}\label{f1}
\end{figure}

\begin{figure}[!htb]
\centering
\includegraphics[height=2.5in, width=3.3in]{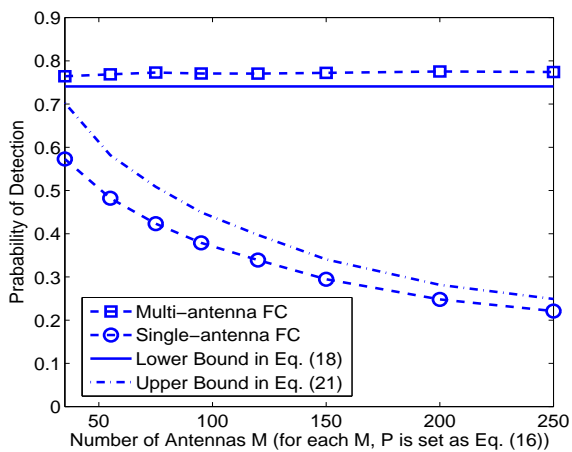}
\caption{Probability of detection vs. number of antennas $M$.}\label{f2}
\end{figure}

In Fig.~\ref{f1}, $M=50$ and we compare the detection performance of the single- and multi-antenna FC under different transmit gain constraints $P$.  When $P$ is small (around $0.1$), we observe that $P_{D}$ is twice that of $P_{D}^s$, and as $P$ increases, both $P_{D}$ and $P_{D}^s$ converge to the same upper bound predicted by~(\ref{eq:ubm}) and~(\ref{eq:ubsh}).  The convergence of $P_D$ to the bound is significantly faster than for $P_D^s$.  Fig. \ref{f2} shows detection performance as a function of $M$, assuming that the sensor transmit gains are reduced as $M$ increases according to $P=\frac{1}{2M}\sum_{i=1}^N \frac{\sigma_n^2d_i^{2\alpha}}{\sigma_{v,i}^2}$.  The performance of the single-antenna FC is also plotted assuming the same decrease in $P$ according to $M$.  As predicted, the detection probability for the multi-antenna FC is constant as $M$ increases and $P$ correspondingly decreases, and is close to the lower bound of~(\ref{eq:lbm}).  However, the performance of the single-antenna FC degrades with an equivalent decrease in $P$, approaching the upper bound in~(\ref{eq:ubsl}) as $P\to 0$. 

\section{Conclusion}
In this paper, we considered the problem of detection in an analog wireless sensor network with a fusion center (FC) possessing a large number of antennas.  An optimization problem was formulated to choose the sensor transmission gains in order to minimize the ratio of the log probability of detection to the log probability of false alarm, and a closed-form expression for the solution was found.  It was shown that a decrease in sensor transmit power can be compensated for by a corresponding increase in the number of FC antennas, asymptotically leading to constant detection performance for a fixed false alarm rate.  Upper and lower performance bounds were also derived for both single- and multi-antenna scenarios.  The benefit of multiple antennas is most pronounced for low transmit powers; at high power, the single- and multi-antenna cases asymptotically converge to the same probability of detection, although the rate of convergence is faster with multiple antennas. To achieve the benefit of the large scale antennas, it requires the FC to have perfect knowledge of the channel state information of all the sensor nodes, which is challenging for the fast fading wireless channels. Future work will include the analysis of energy detector which does not require the knowledge of channel information.

\bibliographystyle{IEEEtran}

\bibliography{reference}

\begin{thebibliography}{10}
\providecommand{\url}[1]{#1}
\csname url@samestyle\endcsname
\providecommand{\newblock}{\relax}
\providecommand{\bibinfo}[2]{#2}
\providecommand{\BIBentrySTDinterwordspacing}{\spaceskip=0pt\relax}
\providecommand{\BIBentryALTinterwordstretchfactor}{4}
\providecommand{\BIBentryALTinterwordspacing}{\spaceskip=\fontdimen2\font plus
\BIBentryALTinterwordstretchfactor\fontdimen3\font minus
  \fontdimen4\font\relax}
\providecommand{\BIBforeignlanguage}[2]{{%
\expandafter\ifx\csname l@#1\endcsname\relax
\typeout{** WARNING: IEEEtran.bst: No hyphenation pattern has been}%
\typeout{** loaded for the language `#1'. Using the pattern for}%
\typeout{** the default language instead.}%
\else
\language=\csname l@#1\endcsname
\fi
#2}}
\providecommand{\BIBdecl}{\relax}
\BIBdecl

\bibitem{Chamberland:2004}
J.-F. Chamberland and V.~V. Veeravalli, ``Asymptotic results for decentralized
  detection in power constrained wireless sensor networks,'' \emph{IEEE J. Sel.
  Areas Commun.}, vol.~22, no.~6, pp. 1007--1015, Aug. 2004.

\bibitem{Cui:2007}
S.~Cui, J.-J. Xiao, A.~J. Goldsmith, Z.-Q. Luo, and H.~V. Poor, ``Estimation
  diversity and energy efficiency in distributed sensing,'' \emph{IEEE Trans.
  Signal Process.}, vol.~55, no.~9, pp. 4683--4695, Sep. 2007.

\bibitem{Quan:2010}
Z.~Quan, W.-K. Ma, S.~Cui, and A.~H. Sayed, ``Optimal linear fusion for
  distributed detection via semidefinite programming,'' \emph{IEEE Trans.
  Signal Process.}, vol.~58, no.~4, pp. 2431--2436, Apr. 2010.

\bibitem{Jiang:2013}
F.~Jiang, J.~Chen, and A.~L. Swindlehurst, ``Optimal power allocation for
  parameter tracking in a distributed amplify-and-forward sensor network,''
  \emph{IEEE Trans. Signal Process. (to appear)}.

\bibitem{Smith:2009}
A.~D. Smith, M.~K. Banavar, C.~Tepedelenlioglu, and A.~Spanias, ``Distributed
  estimation over fading \protect{MACs} with multiple antennas at the fusion
  center,'' in \emph{Proc. Asilomar Conf. Signals, Syst. and Comput. 2009},
  Nov. 2009, pp. 424--428.

\bibitem{Banavar:2012}
M.~K. Banavar, A.~D. Smith, C.~Tepedelenlioglu, and A.~Spanias, ``On the
  effectiveness of multiple antennas in distributed detection over fading
  \protect{MACs},'' \emph{IEEE Trans. Signal Process.}, vol.~11, no.~5, pp.
  1744--1752, May 2012.

\bibitem{Jiang:20122}
F.~Jiang, J.~Chen, and A.~L. Swindlehurst, ``Estimation in phase-shift and
  forward wireless sensor networks,'' \emph{IEEE Trans. Signal Process.},
  vol.~61, no.~15, pp. 3840--3851, Aug. 2013.

\bibitem{Marzetta:2010}
T.~Marzetta, ``Noncooperative cellular wireless with unlimited numbers of base
  station antennas,'' \emph{IEEE Trans. Wireless Commun.}, vol.~9, no.~11, pp.
  3590--3600, Nov. 2010.

\bibitem{Hoydis:2013}
J.~Hoydis, S.~ten Brink, and M.~Debbah, ``Massive \protect{MIMO} in the ul/dl
  of cellular networks: How many antennas do we need?'' in \emph{IEEE J. Sel.
  Areas Commun.}, Feb. 2013, pp. 160--171.

\bibitem{Ngo:2013}
H.~Q. Ngo, E.~G. Larsson, and T.~L. Marzetta, ``Energy ans spectral efficiency
  of very large multiuser \protect{MIMO} systems,'' \emph{IEEE Trans. Commun.},
  vol.~61, no.~4, pp. 1436--1449, Apr. 2013.

\bibitem{Kay:19932}
S.~M. Kay, \emph{Fundamentals of Statistical Signal Processing: Detection
  Theory}.\hskip 1em plus 0.5em minus 0.4em\relax NJ: Prentice Hall, 1993.

\bibitem{Boyd:2004}
S.~Boyd and L.~Vandenberghe, \emph{Convex Optimization}.\hskip 1em plus 0.5em
  minus 0.4em\relax Cambridge, U.K.: Cambridge University Press, 2004.

\bibitem{Jiang:20132}
F.~Jiang, J.~Chen, and A.~L. Swindlehurst, ``Wireless sensing with massive
  \protect{MIMO},'' \emph{to be submitted,
  https://webfiles.uci.edu/fjiang1/www/.}

\end{thebibliography}
\end{document}